\documentclass[a4,12pt]{article}

\usepackage{epsf}
\usepackage{amsfonts,amsmath,amssymb}
\newcommand{\bea}{\begin{eqnarray}}
\newcommand{\eea}{\end{eqnarray}}
\def\be{\begin{equation}}
\def\ee{\end{equation}}
\def\ba{\begin{eqnarray}}
\def\ea{\end{eqnarray}}
\def\nn{\nonumber \\}

\newcommand{\LL}{{\cal L}}

\renewcommand{\theequation}{\thesection.\arabic{equation}}

\newcommand{\no}{\nonumber \\}

\def\th{\theta}
\def\a{\alpha}

\begin{document}
\vfill \setcounter{page}{0} \setcounter{footnote}{0}

\title{CFT Duals for Attractor Horizons}
\author{Dumitru Astefanesei,$^1$\thanks{%
E-mail: {\tt dumitru@aei.mpg.de}}~~ Yogesh K. Srivastava,$^{2}$%
\thanks{%
E-mail: {\tt yogesh@mri.ernet.in}}  \\
%EndAName
$^{1} ${\small {Max-Planck-Institut f\"ur Gravitationsphysik,
Albert-Einstein-Institut, 14476 Golm, Germany}}\\
$^{2}${\small {Harish-Chandra Research Institute, Chhatnag Road, Jhusi,
  Allahabad 211 019, India}}}

\maketitle

\begin{abstract}
In this paper we generalize the results of \cite{Guica:2008mu} to 
$5$-dimensional Anti-de Sitter gravity theories with neutral scalars non-minimally 
coupled to gauge fields. Due to the attractor mechanism, the near horizon geometry 
of extremal black holes is universal and is determined by only the charge parameters. 
In particular, we study a class of near horizon geometries that contain 
an $AdS_2\times S^2$ factor after Kaluza-Klein reduction.
In this way we obtain the microscopic entropy of Gutowski-Reall black hole. We also 
point out a possible connection with the $AdS_2/CFT_1$ correspondence.
\end{abstract}

%\preprint{TIT/HEP-576}

\thispagestyle{empty}

\newpage

\section{Introduction}
Recently, the Kerr/CFT correspondence \cite{Guica:2008mu} has been used 
extensively to understand the statistical entropy of stationary extremal 
black holes. These studies are based on the universality character of the 
near horizon geometry of extremal black holes. More precisely, the isometry 
group of the near horizon geometry is enhanced to $SO(2,1)\times U(1)^{d-3}$ 
in $d=4,5$ dimensions \cite{Bardeen:1999px, Kunduri:2007vf, Astefanesei:2007bf}.
Thus, the near-horizon states of an extremal black hole could be identified 
with a certain two-dimensional chiral conformal field theory 
\cite{Guica:2008mu,Azeyanagi:2008kb,Chow:2008dp,Lu,Loran:2009cr,toate}.
\footnote{In fact, the isometry of the near horizon geometry is 
$SL(2,R)_R \times U(1)_L$ and so the right movers are in the 
ground state. The zero mode of Virasoro algebra of the Kerr/CFT 
correspondece generates $U(1)_L$ --- since the zero mode of 
$SL(2,R)_R$ is $\partial_t$, the right movers are related to 
the entropy away from extremality.}

The analysis in \cite{Guica:2008mu} is similar with the one proposed 
by Brown and Henneaux for $AdS_3$ \cite{Brown:1986nw}. In the Hamiltonian 
formalism , the global charges appear as the canonical generators of the 
asymptotic symmetries of the theory. For each such infinitesimal symmetry, 
there is an associated phase space function that generates the corresponding 
 transformation of the canonical variables.

The asymptotic 
conditions in \cite{Brown:1986nw} are the most general for $AdS_3$ Einstein 
gravity and they respect the following important consistency requirements 
\cite{Henneaux:1985tv}: they are invariant under the $AdS$ group; they 
decay sufficiently slowly to the exact $AdS$ so that to contain the 
spinning black holes; the fall-off is sufficiently fast so that the 
conserved charges are finite. It is also important to emphasize that the 
asymptotic behaviour of the metric in the presence of matter fields can 
be different from that arising from pure gravity. Consequently, the standard 
asymptotic conditions should be relaxed. However, it was shown 
(see, e.g., \cite{Henneaux:2009pw}) that the boundary conditions can be relaxed so that 
the original symmetry is still preserved --- though, the charges are 
modified in order to take into account the presence of the matter fields.

Obviously, if the theory is slightly modified, the boundary conditions
should also be modified in order to accomodate the new solutions of 
physical interest. In \cite{Guica:2008mu} the near horizon geometry involves 
a fibration over $AdS_2$ and so it is another phase space of extremal horizons 
with a different set of boundary conditions. That is some of the deviation 
metric $(h_{\mu\nu})$ components are at the same order in inverse powers of 
$r$ as the corresponding components in the background metric itself. However, 
these boundary conditions still yield finite charges and give rise to a Virasoro 
algebra. The construction of phase spaces containing arbitrary functions 
in the leading components of the metric has been done before [1] (see, e.g.,
\cite{Compere:2007in}).

In this paper we consider extremal stationary black holes in Einstein gravity 
coupled to abelian gauge fields and neutral scalars. Due to the enhanced symmetry 
of the near horizon geometry, the attractor mechanism \cite{Ferrara:1995ih}
can be extended to general extremal spinning black holes 
\cite{Astefanesei:2006dd}. Unlike the non-extremal case 
for which the near horizon geometry (and the entropy) depends on
the values of the moduli at infinity, in the extremal case, 
the near horizon geometry is
universal and is determined by only the charge parameters. This 
is interpreted as a signal that a clear connection to the microscopic 
theory is possible.

We discuss in detail the attractor mechanism for a class of near 
horizon geometries that become $AdS_2\times S^2$ after Kaluza-Klein 
(KK) reduction. We use the entropy function formalism \cite{Sen:2005wa, 
Astefanesei:2006dd, Sen:2007qy} to explicitly show that the 
entropy is independent of the asymptotic values of the
scalars.

Thus, based on these observations, we argue that the Kerr/CFT correspondence
can be generalized to a large class of black holes. A particular 
example of great interest is the Gutowski-Reall (GR)
black hole \cite{Gutowski:2004ez} for which an understanding of the 
statistical entropy is lacking. Our emphasis is mainly on understanding 
the relationship between Kerr/CFT correspondence and $AdS_2/CFT_1$ duality.

In five dimensions there are two distinct asymptotic Virasoro algebras 
\cite{Azeyanagi:2008kb,Chow:2008dp,Lu} that can be obtained by imposing 
appropriate boundary conditions. Even if the corresponding central 
charges are different, the statistical entropies computed by using the 
Cardy formula are equal and match the Bekenstein-Hawking entropy. Since 
these algebras act on the Hilbert states of the CFT, it seems that there 
exist two distinct holographic duals. 

Inspired by the proposal of \cite{sen}, we compute the central charge 
in the $AdS_2$ geometry obtained by KK reduction 
of GR black hole to two dimensions. Interestingly enough, we found that 
it is proportional to the entropy and this may be a hint that there is a 
connection between the Kerr/CFT correspondence and 
the $AdS_2/CFT_1$ duality. However, at this point, it is not clear 
to us if this is indeed the case.
 
The paper is organised as follows: in section 2 we argue that for
extending the Kerr/CFT correspondence to more general theories 
with massless scalars and gauge fields the attractor mechanism 
plays a crucial role. In section 3, we present a concrete analysis 
of the entropy function for a class of near horizon geometries which 
contain an $AdS_2\times S^2$ factor after KK reduction. In section 
4 we show that GR black hole belongs to this class and apply the 
Kerr/CFT correspondence to compute its statistical entropy. In section 
5 we present an analysis of the near horizon geometry of GR black 
hole from a two dimensional point of view. This analysis suggets 
a possible connection with the $AdS_2/CFT_1$ duality. Finally, we 
end with a discussion of our results in section 6.

\section{Attractor mechanism}
In this section we discuss  the attractor 
mechanism for extremal spinning black holes in AdS. Based on the results of 
\cite{Astefanesei:2006dd} we argue on general grounds that there is an attractor 
mechanism for extremal stationary black holes in AdS. 

It is now well understood that supersymmetry does not really play a fundamental role in
the attractor phenomenon. The attractor mechanism works as a consequence of the
$SO(2,1)$ symmetry of the near horizon extremal geometry. This symmetry arises because 
the near horizon geometry involves a fibration over $AdS_2$. The infinite throat 
of $AdS_2$ is at the basis of the attractor mechanism (see \cite{Kallosh:2006bt} 
and section $(4.3)$ of \cite{Astefanesei:2006sy} for a detailed discussion on 
the physical interpretations). Therefore, 
the scalars vary radially, but they are `attracted' to fixed values at the horizon 
(if the entropy function does not have flat directions)
depending only on the charge parameters --- for the stationary black holes the values of 
the scalars at the horizon have also an angular dependence.

For the application of Kerr/CFT analysis, the attractor mechanism 
is crucial. Since the Kerr/CFT analysis is done in the near horizon 
limit and it is usually difficult to extend the notion of Frolov-Thorne 
(FT) vacuum \cite{Frolov:1989jh} all the way to asymptotic infinity, it 
is crucial that the analysis does not depend on asymptotic values of the moduli.

We consider a theory of gravity coupled to a set of masless scalars and vector fields, whose
general action has the form\footnote{In $D=5$ it is possible
to include an additional `$AFF$' Chern-Simons (CS) term.}
\begin{eqnarray}
  \nonumber
  I[G_{\mu\nu},\phi^i,A_{\mu}^I]
  \!&=&\!
  \frac{1}{k^2}\int_{M} d^{5}x \sqrt{-G}[ R-2g_{ij}(\phi)\partial_\mu\phi^i\partial^\mu\phi^j-f_{AB}(\phi)F^{A}_{\mu\nu}F^{B\, \mu\nu} \\
  &&-\frac{1}{2\sqrt{-G}}g_{ABC}(\phi) F^A_{\mu \nu}
  F^B_{\rho \sigma} A^C_{\nu} \epsilon^{\mu \nu \rho \sigma \nu} +V(\phi^i)]\, 
  \label{actiongen}
\end{eqnarray}
where $F^B=dA^B$  with $B=(0, \cdots N)$ are the gauge fields, $\phi^i$ with
($i=1, \cdots, n$) are the scalar fields, and $k^2=16\pi G_5$. We use Gaussian 
units to avoid extraneous factors of $4\pi$ in the gauge fields, and the Newtons's 
constant is set to $G_5=1$. This action resembles that of the {\it gauged} supergravity 
theories.\footnote{The {\it gauged} 
supergravity theories contain a potential for the scalar fields. When 
there are no scalar fields the distinction between gauged and ungaged theories 
is made by the cosmological constant.}

We are interested in stationary black hole solutions of the equations of motion. In
general relativity the boundary conditions are fixed. However, in string theory one can
obtain interesting situations by varying the asymptotic values of the moduli and so, in
general, the asymptotic moduli data should play an important role in characterizing these
solutions. However, due to the enhanced symmetry $SO(2,1)\times U(1)^{d-3}$
of the near horizon geometry of extremal black holes the entropy is independent of 
asymptotic data.

To study the attractor mechanism of these solutions we use the entropy 
function formalism of \cite{Astefanesei:2006dd}.\footnote{Entropy function 
formalism was applied to black holes in AdS space in \cite{ Morales:2006gm}.} 
However, the existence 
of a Chern-Simons term in the action is problematic --- the entropy 
function method relies on gauge as well as diffeomorphism invariance 
of the Lagrangian density. The apparent lack of 
gauge invariance is usually tackled via a 4D reduction \cite{Sahoo:2006vz, Sen:2007qy}
(though, see \cite{Arsiwalla:2008gc}). 

The most general field configuration consistent with the symmetry of the near horizon 
geometry of an extremal spinning black hole is of the form \cite{Astefanesei:2006dd} 
\ba
\label{met}
ds^2 &=& v_1(\th) \left( -r^2 dt^2 + \frac{dr^2}{r^2}\right) 
+\beta(\th) d\th^2 
+ M_{ab}(\th) 
(d\phi^a + \a^a r dt) (d\phi^b + \a^b r dt)
\\ 
\label{gau}
A^M &=& e^M r dt + b^M_a(\th) (d\phi^a + \a^a r dt)
\\
\label{sca}
\phi^s &=& u^s(\th)
\ea
where $\a^a$ and $e^M$ are constants, and
$v_1, v_2, u^s,$ and $\beta$ are functions of $\th$. The form of 
the metric implies that the black hole has zero temperature.

At this point it is important to emphasize the existence of 
two distinct branches of stationary extremal black hole solutions which, in \cite{Astefanesei:2006dd},
are dubbed `ergo-' and `ergo-free' branches according to their properties. 
The first branch,
also known as the fast branch, can exist for angular momentum of magnitude larger than a
certain lower bound and does have an ergo-region. On the other hand, the ergo-free branch
can exist only for angular momentum of magnitude less than a certain upper bound. The
ergo-free branch can be smoothly connected to a static extremal black hole.

Interestingly enough, for the ergo-branch, 
despite (some of) the near horizon scalar fields 
being dependent of the asymptotic data, the entropy is independent of the scalars.
Thus, one can still apply the Kerr/CFT correspondence in this case.

\section{Entropy function}
We discuss in detail the 
entropy function formalism for the most general geometry that has an 
$AdS_2\times S^2$ after KK reduction --- a particular case is GR black hole.
 
In what follows, we are interested in the most general metric that has an 
$AdS_2\times S^2$ after KK reduction:
\be
\label{kk}
ds^2=g_{\alpha\beta}dx^{\alpha}dx^{\beta}=G_{ab}dx^adx^b+u^2(d\phi+\bar{A}_adx^a)^2
\ee
where
\be
G_{ab}dx^adx^b=v_1(- r^2 dt^2 + r^{-2} dr^2)+v_2(d\theta^2+
\sin^2\theta d\psi^2)
\ee

After KK reduction, the KK gauge field appears as a gauge field in four dimensions. 
In order to apply the entropy function method, one should also consider a KK gauge potential 
that respects the symmetry of $AdS_2\times S^2$. We are interested in a KK gauge potential with the following components: $\bar{A}_\theta=0$, $\bar{A}_\psi=\bar{p}\cos\theta$, and the other two components are functions of $r$. The gauge field also preserves the symmetries of the near-horizon geometry and so the gauge potential is given by
\be
A=A_\alpha dx^\alpha=erdt + p\cos\theta d\psi + b\left[ d\phi+A_r(r) dr\right]
\ee

Thus, the KK and original field configurations in four 
dimensions are given by:
\bea
\bar{F}&=&\frac{1}{2}\bar{F}_{\mu\nu}\,dx^\mu\wedge dx^\nu=\bar{e}\,dr\wedge dt-\bar{p}\sin\theta\, d\theta\wedge d\psi
\nn
F&=&\frac{1}{2}F_{\mu\nu}\,dx^\mu\wedge dx^\nu=e\,dr\wedge dt-p\sin\theta\, d\theta\wedge d\psi
\eea   

We use the following results of the dimensional reduction
\bea
g_{\alpha\beta}dx^{\alpha}dx^{\beta}&=&G_{ab}dx^adx^b+G_{AB}(dy^A+\bar{A}_a^Adx^a)(dy^B+\bar{A}_a^Bdx^a)
\nn
\sqrt{-g}&=&\sqrt{-G}\sqrt{\det(G_{AB})}
\eea
\bea
R_5&=&R_4-\frac{1}{4}G^{ac}G^{bd}G_{AB}F^{A}_{ab}F^{B}_{cd}+\frac{1}{4}\partial_aG_{AB}\partial^aG^{AB}-\frac{1}{4}G^{AB}\partial_{a}G_{AB}G^{CD}\partial^{a}G_{CD}-
\nn
& & -\partial_a(G_{AB}\partial^aG_{AB})
\eea
to rewrite the 4-dimensional action in the near-horizon limit (the scalars are constant) 
as :
\be
S_4=\frac{1}{(k_4)^2}\int d^4x \left[u\sqrt{-G}(R_4-\frac{1}{4}u^2\bar{F}^2 - F^2 + \frac{12}{\ell^2}) -
\frac{2A_\phi}{\sqrt{3}}\, F_{tr}\, F_{\psi\theta}\right]
\ee

The quantities $u,b,v_1,v_2,e,\bar{e},p,$ and $\bar{p}$ are constants labelling the
background. We now define:
\ba
%\label{e1}
f(\vec u, \vec v, \vec e, \vec p)\equiv\int d\theta \, 
d\phi\, \sqrt{-G} \, \LL\,
\ea
evaluated for this background. Furthermore, the definitions for the charges and 
the entropy function are 
\be 
\label{e2}
{ q\equiv \frac{\partial f} {\partial e} } \qquad { \bar{q}\equiv \frac{\partial f}{\partial \bar{e}} } \qquad 
E\equiv 2\pi\left[eq+\bar{e}\bar{q}-f(u,b,v_1,v_2,e,\bar{e},p,\bar{p})\right]
\ee
so that $E/2\pi$ is the Legendre transform of 
the function $f$ with respect to
the variables $\{e, \bar{e}\}$.
Thus it follows as a consequence of the equations of motion that, for a
black hole carrying electric 
charge $\vec q=(q,\bar{q})$ and magnetic charge $\vec p = (p,\bar{p})$,
the constants
$\vec v = (v_1, v_2)$, $\vec u = (u, b)$ and $\vec e = (e, \bar{e})$ are given by:
\be 
\label{e3}
{ {\partial E \over \partial u}=0 \qquad 
 {\partial E \over \partial b}=0 \qquad
{\partial E \over \partial v_1}=0\, \qquad  {\partial
E \over \partial v_2}=0}\, 
\ee
\be \label{e4}
e = {1\over 2\pi} \, {\partial E(\vec u, \vec v, \vec q, \vec p) 
\over \partial q}  \qquad \bar{e} = {1\over 2\pi} \, {\partial E(\vec u, \vec v, \vec q, \vec p) 
\over \partial{\bar{q}}}
\ee
Then, the entropy associated with the black hole is given by
\be \label{e5}
S_{BH} = E(\vec u, \vec v, \vec q, \vec p) \, 
\ee
evaluated at the extremum (\ref{e3}).

A straightforward calculation gives

\bea 
\label{centc}
f(\vec u, \vec v, \vec e, \vec p) &=& \frac{4\pi}{(k_4)^2} v_1v_2u\left[-\frac{2}{v_1}+
\frac{2}{v_2}-\frac{1}{2}u^2\left(-\frac{\bar{e}^2}{v_1^2}+\frac{\bar{p}^2}{v_2^2}
\right)-2\left(-\frac{e^2}{v_1^2}+\frac{p^2}{v_2^2}
\right)+\frac{12}{l^2}\right]-
\nn
&+&\frac{4\pi}{(k_4)^2}\frac{16ep\,b}{\sqrt{3}} 
\eea

By combining the equations for $v_1$ and $v_2$ we obtain the following relation

\be
-\frac{2}{v_2}+\frac{2}{v_1}-\frac{24}{\ell^2}=0
\ee

Unlike the theory of gravity with two derivatives in flat space case where the two radii
are equal, in the $AdS$ space the radii are different (see, e.g., GR black hole). Using 
the attractor equations we can rewrite (\ref{centc}) as
\be
f=-\frac{4\pi}{(k_4)^2}\frac{uv_2}{v_1}(-u^2\bar{e}^2+2v_1-4e^2)+\frac{4\pi}{(k_4)^2}
\frac{16epb}{\sqrt{3}}
\ee
and we obtain the entropy $S=16\pi^2 uv_2/(k_4)^2=32\pi^3 uv_2/(k_5)^2=2\pi^2uv_2$.

The entropy function formalism can also be extended to black holes with an $AdS_3$ 
factor in the near-horizon geometry by using the following relation 
between $AdS_3$ and $AdS_2$ metrics:
\be
\label{ads3}
ds_3^2 = v_1 (- r^2 dt^2 + r^{-2} dr^2) +
u^2(d\phi + \bar{A}rdt)^2  
\ee 
where the constraint $v_1=(u\bar{A})^2$ assures that the geometry 
(\ref{ads3}) is $AdS_3$. 

\section{Gutowski-Reall black hole and its near horizon geometry}
In what follows we recapitulate the main results of 
\cite{Reall:2002bh, Gutowski:2004ez} and rewrite the near-horizon 
geometry in a form suitable to our analysis. We explicitly show 
that, indeed, there is an $AdS_2$ in the near horizon geometry of 
GR black hole and obtain the KK reduction to four dimensions. Finally,
we use the Kerr/CFT correspondence to compute the statistical 
entropy of GR black hole.

\subsection{Generalities}
The theory we shall be considering is minimal $D=5$ {\it gauged}
supergravity with bosonic action
\bea
 S_5 &=& \frac{1}{4\pi G_5}\int \left[\left( \frac{R_5}{4} + \frac{3}{\ell^2}
 \right) \star 1 - \frac{1}{2} F \wedge \star F - \frac{2}{3\sqrt{3}} F
 \wedge F \wedge A \right]\nn
&=&{\frac{1}{(k_5)^2}}\int d^5x \left[\sqrt{-g}(R_5 -F^2+\frac{12}{\ell^2}) -
{\frac{2}{3\sqrt{3}}}\varepsilon^{\alpha\beta\gamma\tau\delta}
A_\alpha F_{\beta\gamma}F_{\tau\delta}\right]
\eea
where $R_5$ is the Ricci scalar, ${}F^2\equiv
{}F_{\alpha\beta}{}F^{\alpha\beta}$, and
$F=dA$ is the field strength of the $U(1)$ gauge field. We also use the notation 
$(k_D)^2=16\pi G_D$, where $G_D$ is the gravitational constant in $D$ dimensions.
The bosonic equations of motion are
\bea
\label{eqofmot}
{}^5 R_{\alpha\beta}-2{}F_{\alpha\gamma}{}F_\beta{}^\gamma
+\frac{1}{3}g_{\alpha\beta}{}(F^2+{\frac{12}{\ell^2}})&=&0\nn d*{}F +
\frac{2}{\sqrt{3}} {}F \wedge {}F&=&0
\eea

In flat space \cite{Reall:2002bh}, the geometry of the event horizon of 
any supersymmetric black hole of minimal $5$-dimensional supergravity must 
be $T^3$, $S^1 \times S^2$, or a quotient of a homogeneously squashed $S^3$.  
However, there is no general classification of the near horizon 
geometries of susy black holes in AdS spacetime.

In AdS space \cite{Gutowski:2004ez},  Gutowski and Reall 
found an interesting solution that is asymptotically AdS and does not 
have an $AdS_3$ component in the near-horizon geometry. In the ungauged 
theory the near-horizon geometry of a BPS black hole is maximally supersymmetric. In 
the gauged supergravity this is not true because the only maximally supersymmetric 
solution is $AdS_5$. 

The ansatz for the full metric in Guassian coordinates is \cite{Reall:2002bh}
\be
\label{eqnmetric}
 ds^2 = -r^2 \Delta^2 dU^2 + 2 dU dr + 2 r h_A dU dx^A + \gamma_{AB} dx^A dx^B
\ee
where $\gamma_{AB}$ is a function of $r$ and $x^A$. This metric guarantees the 
existence of a regular near horizon geometry, defined by the limit $r = \epsilon
\tilde{r}$, $U = \tilde{U}/\epsilon$ and $\epsilon \rightarrow 0$. 

The horizon, $r=0$, is a Killing horizon of $V=\partial /\partial U$ --- the 
near-horizon metric has the same form (\ref{eqnmetric}), but with $\Delta$, $h_A$, 
and $\gamma_{AB}$ depending only on $x^A$. The gauge field $A$ in the near-horizon 
limit (${\cal L}_V A = 0$) is given by
\be
 A = \frac{\sqrt{3}}{2} r \Delta dU + a_A dx^A
\ee

\subsection{The near horizon geometry}

For $\Delta > \sqrt{3}/\ell$ the near-horizon solution is
\ba
\label{eqn:sol3}
 ds^2 &=& - r^2 \Delta^2 dU^2 + 2 dU dr - \frac{6\Delta r}{\ell (\Delta^2
- 3 \ell^{-2})} dU (d \phi + \cos \theta d\psi ) \nn
   &+& \frac{1}{\Delta^2 - 3 \ell^{-2}} \left[ \frac{\Delta^2}{\Delta^2 -
3 \ell^{-2}} \left( d\phi + \cos \theta d \psi \right)^2 + d\theta^2 +
\sin^2 \theta d\psi^2 \right] \nn
 F &=& - \frac{\sqrt{3}}{2} \Delta dU \wedge dr + \frac{\sqrt{3} \sin
\theta}{ 2 \ell (  \Delta^2 - 3 \ell^{-2} ) } d \theta \wedge d\psi
\ea
where $\Delta$ is constant everywhere. 

Dimensional reduction on $\partial/\partial \phi$ yields an $AdS_2 \times S^2$ 
geometry. We rewrite (\ref{eqn:sol3}) in a suitable form for KK reduction:
\ba
\label{misto}
ds^2 &=& - r^2 \Delta^2 dU^2 + 2 dU dr +  
\left(\frac{\Delta}{\Delta^2 - 3 \ell^{-2}}\right)^2 \left[d\phi + \cos \theta d \psi 
 - \frac{3r}{\ell\Delta}(\Delta^2 - 3 \ell^{-2})dU \right]^2-\nn
   &-& \frac{9r^2}{\ell^2}dU^2+\frac{1}{\Delta^2 - 3 \ell^{-2}}(d\theta^2 +
\sin^2 \theta d\psi^2)
\ea
To make the $AdS_2$ part manifest, we introduce a new coordinate
\be
\tau=(\Delta^2 + 9 \ell^{-2})U+\frac{1}{r} \qquad d\tau=(\Delta^2 + 9 \ell^{-2})dU-\frac{dr}{r^2}
\ee
and rewrite (\ref{misto}) as
\ba
\label{GR}
ds^2 &=&\frac{1}{\Delta^2 + 9 \ell^{-2}}(-r^2d\tau^2+r^{-2}dr^2) +\frac{1}{\Delta^2 - 3 \ell^{-2}}(d\theta^2 +
\sin^2 \theta d\psi^2) 
\nn
&+&\left(\frac{\Delta}{\Delta^2 - 3 \ell^{-2}}\right)^2 \left[d\phi + \cos \theta d \psi 
 - \frac{3r}{\ell\Delta}\frac{\Delta^2 - 3 \ell^{-2}}{\Delta^2 + 9 \ell^{-2}}(d\tau+\frac{dr}{r^2}) \right]^2
\ea

\subsection{Boundary conditions and central charges}

Let us consider a perturbation of the near horizon metric 
($g_{\mu\nu}$). If $h_{\mu\nu}$ is some deviation from it 
the new metric is given by $\bar{g}_{\mu\nu}=g_{\mu\nu}+h_{\mu\nu}$. 
Following \cite{Azeyanagi:2008kb, Lu} we see that, being in five 
dimensions, we can have two
consistent boundary conditions corresponding to two $U(1)$s such that the 
diffeomorphisms will generate two copies of chiral Virasoro algebra.

One of the possible boundary conditions for $h_{\mu\nu}$ is
\begin{align}
\left(
\begin{array}{ccccc}
h_{\tau\tau}=\mathcal{O}(r^2)
&h_{\tau r}=\mathcal{O}(\frac{1}{r^2})
&h_{\tau \theta}=\mathcal{O}(\frac{1}{r})
&h_{t\psi}=\mathcal{O}(r)
&h_{t\phi}=\mathcal{O}(1)\\
h_{r\tau}=h_{\tau r}
&h_{rr}=\mathcal{O}(\frac{1}{r^3})
&h_{r\theta}=\mathcal{O}(\frac{1}{r^2})
&h_{r\psi}=\mathcal{O}(\frac{1}{r^3})
&h_{r\phi}=\mathcal{O}(\frac{1}{r^2}) \\
h_{\theta \tau}=h_{\tau\theta}
&h_{\theta r}=h_{r\theta}
&h_{\theta\theta}=\mathcal{O}(\frac{1}{r})
&h_{\theta\psi}=\mathcal{O}(\frac{1}{r})
&h_{\theta \phi} = \mathcal{O}(\frac{1}{r}) \\
h_{\psi \tau}=h_{\tau \psi}
&h_{\psi r} =h_{r\psi}
&h_{\phi\theta} =h_{\theta\phi}
&h_{\psi\psi}=\mathcal{O}(\frac{1}{r})
&h_{\psi \phi}=\mathcal{O}(1)\\
h_{\phi \tau}=h_{\tau \phi}
&h_{\phi r}= h_{r\phi}
&h_{\phi \theta}=h_{\theta \phi}
&h_{\psi\phi} =h_{\phi \psi}
&h_{\phi\phi}=\mathcal{O}(1)
\end{array}
\right)
\label{ry_boundary_condition}
\end{align}

We give the details about how to get the most general diffeomorphism 
that preserves these boundary conditions in the appendix. We obtain 
that the most general diffeomorphism that preserves 
(\ref{ry_boundary_condition}) is given by
\begin{align}
\zeta &=
\Bigl[C+\mathcal{O}\bigl(\frac{1}{r^3}\bigr)\Bigr]\partial_t
+[-r\gamma'(\phi)+\mathcal{O}(1)]\partial_{r}
+\mathcal{O}\bigl(\frac{1}{r}\bigr)\partial_{\theta} \no
&\quad +\mathcal{O}\bigl(\frac{1}{r^2}\bigr)\partial_{\psi}
+\Bigl[\gamma(\phi)+\mathcal{O}\bigl(\frac{1}{r^2}\bigr)\Bigr]\partial_\phi
\end{align}
where $C$ is an arbitrary constant and
$\gamma(\phi)$ is an arbitrary function of $\phi$.
From this, 
the asymptotic symmetry group is generated by 
the diffeomorphisms of the form 
\begin{align}
\zeta^t&=\partial_{t} \label{time_killing}\\
\zeta^\phi_\gamma&=\gamma(\phi)\partial_{\phi}-r\gamma'(\phi)\partial_r 
\label{ry_generator}
\end{align}
Especially, (\ref{ry_generator})
generates the conformal group of one of the $U(1)$ circles. A generator 
of the Virasoro algebra of the chiral $CFT_2$ is 
identified with this class of diffeomorphisms 
which preserve the  appropriate boundary condition on 
the near horizon geometry.
To see that it really obeys the Virasoro algebra,
we expand $\gamma(\phi)$ in modes and define 
$\gamma_{n}=-e^{-in\phi}$. Then, it can be easily seen that 
$\zeta^{\phi}_{n}$, which are defined as
\begin{align}
\zeta^{\phi}_{n}=\gamma_{n}\partial_{\phi}-
r\gamma_n'\partial_r
\end{align} 
obey the Virasoro algebra under the Lie bracket as 
\begin{align}
[\zeta^{\phi}_m,{\ }\zeta^\phi_n]_{Lie} =-i(m-n)\zeta^{\phi}_{m+n}
\end{align}
We notice that 
the Virasoro generators are constructed from $r$ and $\phi$.
In other words, we see that the generators of the Virasoro algebra
act on only $\phi$-direction in the dual boundary field theory.
Thus it is very different from 
the usual holographic dual $CFT_2$ where the time direction $t$
play some role. It seems that we cannot describe
dynamical processes by using this Virasoro algebra,
but at least to calculate the entropy,
we can use the Virasoro algebra on the $\phi$-direction.

The allowed symmetry transformations include time translations 
generated by $\zeta^{t}$ which correspond to energy above extremality. 
Since we study only extremal black holes, we set the 
corresponding conserved charge $Q_{\partial_{t}}=0$. This restriction 
is consistent because $\zeta^t$ commutes with other generators in 
the asymptotic symmetry group. 

The other allowed boundary condition is
\begin{align}
\left(
\begin{array}{ccccc}
h_{tt}=\mathcal{O}(r^2)
&h_{tr}=\mathcal{O}(\frac{1}{r^2})
&h_{t\theta}=\mathcal{O}(\frac{1}{r})
&h_{t\phi}=\mathcal{O}(1)
&h_{t\phi}=\mathcal{O}(r)\\
h_{rt}=h_{tr}
&h_{rr}=\mathcal{O}(\frac{1}{r^3})
&h_{r\theta}=\mathcal{O}(\frac{1}{r^2})
&h_{r\psi}=\mathcal{O}(\frac{1}{r})
&h_{r\phi}=\mathcal{O}(\frac{1}{r^2}) \\
h_{\theta t}=h_{t\theta}
&h_{\theta r}=h_{r\theta}
&h_{\theta\theta}=\mathcal{O}(\frac{1}{r})
&h_{\theta\psi}=\mathcal{O}(\frac{1}{r})
&h_{\theta \phi} = \mathcal{O}(\frac{1}{r}) \\
h_{\psi t}=h_{t\psi}
&h_{\psi r} =h_{r\psi}
&h_{\psi\theta} =h_{\theta\psi}
&h_{\psi\psi}=\mathcal{O}(1)
&h_{\psi \phi}=\mathcal{O}(1)\\
h_{\phi t}=h_{t\phi}
&h_{\phi r}= h_{r\phi}
&h_{\phi\theta}=h_{\theta \phi}
&h_{\phi\psi} =h_{\psi\phi}
&h_{\psi\psi}=\mathcal{O}(\frac{1}{r})
\end{array}
\right)
\label{rphi_boundary_condition1}
\end{align}
and the general diffeomorphism preserving \eqref{rphi_boundary_condition1} can be
written as
\begin{align}
\zeta &=
\Bigl[C+\mathcal{O}\bigl(\frac{1}{r^3}\bigr)\Bigr]\partial_t
+[-r\epsilon'(\psi)+\mathcal{O}(1)]\partial_{r}
+\mathcal{O}\bigl(\frac{1}{r}\bigr)\partial_{\theta} \no
&\quad +\Bigl[\epsilon(\psi)+\mathcal{O}\bigl(\frac{1}{r^2}\bigr)\Bigr]\partial_{\psi}
+\mathcal{O}\bigl(\frac{1}{r^2}\bigr)\partial_{\phi}
\end{align}
where $C$ is an arbitrary constant and
$\epsilon(\psi)$ is an arbitrary function of $\psi$.
The asymptotic symmetry group (ASG) is generated by $\zeta^t$ and
\begin{align}
\zeta^\psi_\epsilon=\epsilon(\psi)\partial_\psi-r\epsilon'(\psi)\partial_r
\label{rphi_generator}
\end{align}
In exactly the same manner as above,
we define $\epsilon_n =-e^{-in\psi}$ and so
\begin{align}
\zeta^{\psi}_n = \epsilon_{n}\partial_{\psi}-
r\epsilon_n'\partial_r
\end{align}
obey the Virasoro algebra 
\begin{align}
[\zeta^{\psi}_m,{\ }\zeta^{\psi}_n]_{Lie} =-i(m-n)\zeta^{\psi}_{m+n}
\end{align}
In this case
the Virasoro generator is constructed from $r$ and $\psi$.

We will see that these boundary conditions indeed
lead to the correct black hole entropy.\footnote{We would 
also like to point out that we have explicitly checked 
that the contribution of gauge fields and the CS term vanishes. 
Since, recently, this result was proven for the general case --- see 
the note added at the end of the paper --- we do not present the 
details here.} 
As discussed in \cite{Loran:2009cr}, the two CFTs are 
related by a $SL(2,Z)$ modular group transformation that 
interchanges two circles in the near horizon 
geometry and so maps the two CFTs corresponding to 
two circles into each other. 

Following the covariant formalism of the ASG
\cite{Compere}, a conserved charge $Q_{\zeta}$ 
associated with an element $\zeta$ is defined by
\begin{align}
Q_{\zeta} =\frac{1}{8\pi }
\int _{\partial\Sigma} k_{\zeta}[h,g] 
\label{charge_zeta}
\end{align}
where $\partial\Sigma$ is a spatial surface at infinity and
\begin{align}
k_{\zeta}[h,g] &=\frac{1}{4}\epsilon_{\alpha\beta\gamma\mu\nu}
\Bigl[
\zeta^{\nu}D^{\mu}h -\zeta^{\nu}D_{\sigma}h^{\mu\sigma}+
\zeta_{\sigma}D^{\nu}h^{\mu\sigma} \nonumber \\
&\quad +\frac{1}{2}h D^{\nu}\zeta^{\mu}
-h^{\nu\sigma}D_{\sigma}\zeta^{\mu} 
+\frac{1}{2}h^{\sigma\nu}(D^{\mu}\zeta_{\sigma}+D_{\sigma}\zeta^{\mu})
\Bigr]dx^{\alpha}\wedge dx^{\beta}\wedge dx^{\gamma}  
\label{current_zeta}
\end{align}
Here $g_{\mu\nu}$ is the metric of the background geometry 
 and $h_{\mu\nu}$
is deviation from it.
We also notice that the covariant derivative
is defined by using $g_{\mu\nu}$.
In addition to a charge $Q_{\zeta_n}$ associated with $\zeta_n$, 
there exists a charge $Q_{\partial_{\tau}}$ associated with 
$\partial_{\tau}$. As discussed above, this is set to zero to preserve the extremality condition. 

Then let us consider the Dirac bracket of $Q_{\zeta_n}$
under the constraint $Q_{\partial_{\tau}}=0$.
It is determined by considering the transformation property of the charge 
$Q_{\zeta_n}$ under a diffeomorphism generated by $\zeta_m$. 
It then follows that
\begin{align}
\{Q_{\zeta_{m}},Q_{\zeta_n}\}_{Dirac}=Q_{[\zeta_{m},\zeta_{n}]}
+\frac{1}{8\pi }\int_{\partial \Sigma}
k_{\zeta_{m}}[\mathcal{L}_{\zeta_{n}}g, g]
\label{dirac_algebra}
\end{align}
By expanding the charge in terms of $L_{n}$'s 
and replacing the Dirac bracket $\{.,.\}$ 
by the  commutator 
we see that $L_n$ satisfy a Virasoro algebra 
\begin{align}
[L_m,L_n] =(m-n)L_{m+n}+\frac{c}{12}m(m^2 +\alpha)\delta_{m+n,0}
\end{align}

This prescription works for both boundary conditions.
The central charges $c_i$ in these Virasoro algebras, at the level of 
Dirac brackets
of the associated charges $Q^{i}_n$,
can be calculated  
from the $m^3$ terms in the expression
\begin{equation}
\frac{1}{8\pi}\int_{\partial\Sigma} 
k_{\zeta^i_m}[{\mathcal L}_{\zeta^i_{(-m)}}g,g]=
 -\frac{i}{12} (m^3 + \alpha m) c_i
\end{equation}
Using the Lie derivatives calculated in the appendix, we get for 
the first boundary condition (of interest for the next section)
\begin{equation}
c=\frac{36v_1 u}{\ell \Delta}\pi
\end{equation}
To calculate the entropy, we also need to calculate the FT 
temperature \cite{Frolov:1989jh}. Using 
the formula given by Chow et al \cite{toate}, we see that these temperatures 
are given by the constants $k_1$ and $k_2$ appearing in the $dt d\phi$ and 
$dtd\psi$ components of the metric written in the form
\begin{equation}
ds_5^{2} = v_1(-r^2 dt^2 + \frac{dr^2}{r^2}) + v_2 (d\theta^2 + \sin^{2}\theta (e_1- e_2)^2) + 
u^2 \left(e_1 + e_2 + \cos\theta(e_1 -e_2)\right)^2
\end{equation}
where $e_{i} = d\phi_i + k_{i}r dt$ and $\psi= \phi_1 -\phi_2$ and $\phi= \phi_1 + \phi_2$.
In terms of $k_1$ the FT temperatures are given by
\be
k_i=\frac{1}{2\pi T_i} \ , \ \ S= \frac{\pi^2}{3}c_1 T_1 = \frac{\pi^2}{3}c_2T_2
\ee
%%%
So finally we get the following values for the 
entropy\footnote{We have to `trade' the energy for the temperature 
in the usual Cardy formula $S=2\pi\sqrt{cE/6}$ by using the first 
law $dE=TdS$.} and the central charge:
\begin{equation}
S= 2\pi^2 v_2 u \ , \ k=  \frac{3v_1}{\ell \Delta v_2} \ , \ c= \pi\frac{36v_1 u}{\ell \Delta}
\end{equation}
Here $k= \frac{1}{2\pi T_{FT}}$ where $T_{FT}$ is the FT temperature. 

\section{The relation with $AdS_2/CFT_1$}
Since the extremal black holes have an $AdS_2$ in their near horizon 
geometry, it is expected that the dual conformal quantum mechanics (CQM)
living at the boundary plays an important role in understanding their 
statistical entropy. Indeed, it hase been shown in \cite{sen} that 
the entropy function gives rise to an entropy that can be interpreted 
as the logarithm of the ground state degeneracy of the dual CQM in 
a fixed charged sector. Since the CQM is living on the boundary that 
is a circle,  the partition function may be represented as a trace 
over the Hilbert space of the CFT. 

The main result of Sen is a specific relation between degeneracy of 
black holes microstates and an appropriately defined partition function 
of string theory on the near horizon geometry (reffered to as the 
{\it quantum entropy function}). More concretely, the microscopic 
degeneracy $S_{micro}=\ln d_{micro}$ is given by
\be
\label{gigi}
d_{micro}(\vec q)=\langle \exp[-q_M\oint d\theta A_{\theta}^M] \rangle^{finite}_{AdS_2}
\ee
where $\langle \rangle_{AdS_2}$ denotes the unnormalized path integral 
over various fields on Euclidean global $AdS_2$ associated with the 
attractor geometry for charge $\vec{q}$ and $A^M_{\theta}$ are the 
values of gauge fields along the boundary of $AdS_2$. In the classical 
limit this reduces to the usual relation between microscopic entropy and 
macroscopic (Wald) entropy.

In $AdS_2$, the solution to the classical equations of motion for the 
gauge fields has two independent modes near the 
boundary: the constant mode and the mode representing the asymptotic value 
of the electric field. Since the electric field mode is dominant and the 
electric fields determine the charges carried by the black hole, the 
relation (\ref{gigi}) is written for a fixed charge sector. However, one 
can also work with fixed values of the constant modes (a detailed discussion 
can be found in \cite{sen}) and this leads to a new partition function with 
the finite part given by
\be
\label{gigi1}
Z_{AdS_2}^{finite}(\vec{e})=\sum_{\vec{q}}d_{micro}(\vec{q})e^{-2\pi \vec{e}\vec{q}}
\ee
Since we allowed the asymptotic electric fields to fluctuate, the right hand side 
now has a sum over different charges. Due to the fact that this involves integrating over 
non-renormalizable modes, even when such a partition function can 
be defined, it probably only makes sense as an asymptotic expansion around 
the classical limit. However, this is the partition function we are interested 
in.

In the classical limit both (\ref{gigi}) and (\ref{gigi1}) reduce 
to the ususal relation between the statistical and the thermodinamical 
entropies and so the microscopic description of the entropy of 
an extremal black hole for large charges is a direct consequence of 
$AdS_2/CFT_1$ duality in the classical limit.

It is also important to mention a related interesting work of 
Hartman and Strominger \cite{Hartman:2008dq}. This work is especially 
relevant for our discussion. 

In this section, we try to see if there is a relationship between  central charges
calculated using Kerr/CFT correspondence and central charges appearing in 
recent attempts \cite{Hartman:2008dq, Castro:2008ms} to find the central charge 
in $AdS_2$ by applying a Brown-Henneaux procedure. On the face of it, this 
seems unlikely because the vector fields generating the diffeomorphism 
are functions of time in one case $AdS_2$ while they are 
functions of $U(1)$ coordinate in Kerr/CFT analysis. But both of them 
involve modifying the asymptotic boundary conditions. In $AdS_2$ case, 
one needs to twist the energy momentum tensor by a certain $U(1)$ gauge 
transformation while in the Kerr/CFT correspondence one needs to take some 
of the components of the perturbation metric to be of the same order as the 
background.\footnote{We believe that, in fact, in the analysis 
of \cite{Castro:2008ms} these considerations should also be taken in account.}

Let us now discuss GR solutions after KK reduction in two dimensions by using 
the entropy function formalism. By comparing (\ref{GR}) with (\ref{kk}) one 
can read off $v_1, v_2, u,$ as well as the KK gauge potential $\bar{A}_a$. Explicitly, 
we obtain
\begin{equation}
v_1 = \frac{1}{\Delta^2 + 9\ell^{-2}} , \ \  v_2 = \frac{1}{\Delta^2 - 3\ell^{-2}} , \ \ 
u = \frac{\Delta}{\Delta^2 -3\ell^{-2}} = \Delta v_2
\end{equation}
The original gauge potential in five dimensions is
\bea
A=erdt+b(d\phi+\cos\theta d\psi)=
\frac{\sqrt{3}}{2}\Delta\, v_1\, r\, d\tau-\frac{\sqrt{3}}{2l}\, v_2\, 
(d\phi+\cos\theta d\psi)
\eea
and the field strength configurations after KK reduction are are given by
\bea
\bar{F}&=&\frac{1}{2}\bar{F}_{\mu\nu}\,dx^\mu\wedge dx^\nu=-\frac{3}{\ell}
\frac{\Delta^2-3\ell^{-2}}{\Delta(\Delta^2+9\ell^{-2})}\,dr\wedge d\tau-\sin\theta\, d\theta\wedge d\psi
\nn
F&=&\frac{1}{2}F_{\mu\nu}\,dx^\mu\wedge dx^\nu=e\,dr\wedge dt-p\sin\theta\, d\theta\wedge d\psi
\eea
with $e$ and $p$ as in (\ref{eqn:sol3}). From the solution, we get 
\begin{equation}
\overline{e} = -\frac{3 v_1}{\ell\Delta v_2}  , \ \ \overline{p}=1 , \ \ e= \frac{\sqrt{3}\Delta v_1}{2} , \ \  p=b=-\frac{\sqrt{3}v_2}{2\ell}
\end{equation}
In the $AdS_2/CFT_1$ duality, the central charge is given by \cite{Castro:2008ms}
\begin{equation}
c= 3 Vol_l\, \mathcal{L}_{2D}
\end{equation}
where the volume element ${\rm Vol}_l=2\pi l^2$ and Lagrangian
density is related
to the on-shell bulk action by
\begin{equation}
I_{\rm bulk}\big|_{\textrm{\tiny EOM}}=-\int_{\cal M}
d^2x\sqrt{-g}\,{\cal L}_{\rm 2D}~.
\end{equation}
This form of the central charge is consistent with the analysis 
of \cite{sen}. Since 
\ba
\label{e1}
f(\vec u, \vec v, \vec e, \vec p)\equiv\int d\theta \, 
d\phi\, \sqrt{-G} \, \LL\,
\ea
occurs in the entropy function formalism it is worth to compute 
its expression for the GR black hole. It can easily be seen that after 
dimensional reduction, $f(\vec u, \vec v, \vec e, \vec p)$ will correspond 
to the central charge. 

Let us now  evaluate $f(\vec u, \vec v, \vec e, \vec p)$ for GR black hole. 
Replacing the near horizon data in the expression (\ref{centc}) we obtain 
\begin{equation}
f(\vec u, \vec v, \vec e, \vec p) = \frac{4\pi}{k_4^2} v_1 u
\end{equation}
By comparing with th results in the previous section, we tentatively make 
the identification that $f(\vec u, \vec v, \vec e, \vec p)$ is indeed 
proportional to central charge and so the central charges appearing in 
Kerr/CFT and $CFT_1$ are related.\footnote{Similar considerations on 
a relation between central charges in $AdS_2$ and $AdS_3$ in the presence 
of Chern-Simons terms appeared also 
in \cite{Alishahiha:2008rt}, though the Kerr/CFT does not play any role 
in this work.} We are currently 
investigating the possible connection and hope to report on it in near future.

\section{Discussion}

In this paper we propose that the Kerr/CFT correspondence can be applied 
to stationary extremal black holes in gravity theories with massless, neutral scalars non-minimally coupled to gauge 
fields.\footnote{The existence of a {\it non-extremal} black hole 
horizon is considered as a boundary condition in \cite{Solodukhin:1998tc}.} 
Our conclusion 
relies heavily on the existence of the attractor mechanism that fixes 
the entropy of both, ergo and ergo-free, branches independent of the asymptotic data. 

An important observation is that in the case of Kerr/CFT correspondence, 
the Virasoro generators are constructed from $r$ and an angular coordinate 
(e.g., $\phi$). In other words, we see that the generators of the 
Virasoro algebra act on only $\phi$-direction in the dual boundary 
field theory. Thus it is very different from the usual holographic 
dual $CFT_2$ where the time direction $t$ plays some role. It seems 
that we cannot describe dynamical processes by using this Virasoro algebra,
but at least to calculate the entropy, we can use the Virasoro algebra 
on the $\phi$-direction.

The temperature of the dual chiral $CFT_2$ is determined by identifying 
quantum numbers in the near horizon geometry with those in the original 
geometry \cite{Frolov:1989jh}. For spinning black holes, one can give a 
physical interpretation to the rotating spatial coordinates (with the 
horizon's angular velocity). That is they are comoving with the 
radiation fluid environment that is required to equilibrate the black 
hole. 

Frolov and Thorne gave a quantum-field theoretic argument why the 
environment must rotate {\it rigidly}. Local observers which are 
comoving with the fluid environment are the natural observers to 
describe the equlibrium of a system containing a black hole --- they 
see a locally isotropic thermal distribution of quanta \cite{Frolov:1989jh}.
However, it is important to emphasize that these observers are 
not actually suitable for defining global properties of the system. 
Indeed, there is no way for them globally to synchronize their clocks, 
and consequently there is no global time-slicing with respect to 
which they are at rest.

Therefore, for the application of Kerr/CFT analysis, the attractor mechanism 
is crucial. Since the Kerr/CFT analysis is done in the near horizon limit 
and it is difficult to extend the notion of FT vacuum 
all the way to asymptotic infinity, it is crucial that the analysis does 
not depend on asymptotic values of the moduli.

Gravitating systems are never truly isolated in the traditional sense of 
thermodynamics and gravitational `thermodynamics' has to be formulated 
globally because of the infinite range of the gravitational field.

Within the AdS/CFT duality there is a concrete connection between the 
attractor mechanism (gravity side) and the `dual' universality property 
of the QFT \cite{Astefanesei:2007vh} (see also \cite{deBoer:2008ss, Hotta:2008xt}).
The scalars (moduli)
flow has a nice interpretation as an RG flow towards the IR attractor horizon. 
Therefore, the fact (reffered to as `universality' of QFT) that
 the IR end-point of a QFT RG flow does not depend 
upon UV details is equivalent, in the hlography context, to the fact 
that the bulk solution for the small $r$ does not depend upon the 
details of the matter at large values of $r$. Indeed, due to the attractor mechanism 
the black hole horizon (IR region) does not have any memory of the initial 
conditions (the UV values of the moduli) at the boundary. Thus, in the AdS/CFT 
duality context, the Kerr/CFT correspondence has a nice interpretation:
the universality of the near horizon geometry in the IR regime is at the basis 
of the statistical entropy computations that do not depend of details at the boundary. 
This statement is related to the fact that more than one UV quantum field theories 
can flow to the same IR point. 

The existence of the two branches (with and without ergoregion) may be puzzling 
for the Kerr/CFT correspondence. How is it possible that this method is working in 
both cases? The answer is related again to the existence of the attractor mechanism 
\cite{Astefanesei:2006dd}. The entropy function has no flat directions for the ergo-free branch: the scalars and all other
background fields at the horizon are independent of the asymptotic data. However, there is
a drastic change for the ergo-branch --- the entropy function has flat directions:
despite the entropy being independent of the moduli, the near horizon fields are dependent
on the asymptotic data. The existence of an ergo-region allows energy to be extracted
classically either by the Penrose process for point particles or by superradiant
scattering for fields. It is tempting to believe that the presence of the ergo-sphere is
intimately related to the appearance of flat directions. One might say that the
ergo-branch, not completely isolated from its environment due to these processes, retains
some dependence on the asymptotic moduli.  From this perspective, it is amazing that the
black hole is isolated enough for the entropy to remain independent --- however, 
the addition of higher derivative terms might lift these flat directions.

In $AdS$ spacetime there are no static supersymmetric black holes. The extremal limit 
is different than the BPS limit --- in the BPS limit one obtains naked singularities. 
One way to avoid this problem is to construct spinning susy black holes. Gutowski and 
Reall constructed a spinning susy solution in five dimensional minimal 
gauged supergravity.

The main goal of this work was to give an interpretation for the microscopic 
entropy of GR black hole. It is important to mention that, despite the fact that 
this is a susy black hole, a computation of its entropy in the boundary CFT is 
lacking. The attempts to match it with the index (the number of chiral primaries ) 
of the four dimensional CFT failed \cite{Kinney:2005ej}. The reason 
may be that, since the black hole is not maximally supersymmetric, two or more 
short (BPS) multiplets can combine into a long representation. 

As a side observation, we mention that it will be interesting to understand 
the role of the dipole charge of black rings within the Kerr/CFT 
correspondence --- the analysis in \cite{Compere:2007vx} may be useful.\footnote{It 
is known that the dipole charge appears in 
the first law in the same manner as a global charge \cite{eu}.}

To this end, let us comment on a possible relation connection between the 
Kerr/CFT correspondence and the $AdS_2/CFT_1$ duality.
First, note that one can perform a KK reduction to get a two dimensional 
effective theory. For GR black hole all values of the parameters that 
characterize the near horizon geometry are given in section 5. Note that the 
magnetic fields represent flux through the sphere labelled by the angular 
coordinates and should not be explicitly displayed. Thus, one can obtain the 
degeneracy of microstates by using the quantum entropy function proposal of 
Sen \cite{sen}. 

One way to compute conserved charges is by using a canonical realization 
of the ASG. For $AdS_2$ Maxwell-dilaton gravity, Hartman and Strominger 
\cite{Hartman:2008dq} proposed that the usual conformal diffeomorphisms must 
be accompanied by gauge transformations in order to mantain the boundary 
conditions. In this way, the conformal transformations are generated by 
a twisted stress tensor and one can obtain the central charge for $AdS_2$.
Alternatively, one can use a Lagrangian formalism and compute the stress-energy 
tensor for the boundary theory. This method was implemented in \cite{Castro:2008ms}
where it was identified the central charge of $AdS_2$ to be proportional to 
the Lagrangian density in accord with \cite{sen}. However, the meaning of 
the anomalous  
transformation of the stress tensor in the boundary $CFT_1$ is 
not clear, since there is no explicit construction of the $CFT_1$.

Also, as argued in \cite{Azeyanagi:2008dk}, for the case of 
$D1-D5-P$ system with an $AdS_3$ factor in the near horizon limit, 
one can obtain a chiral $CFT_2$ occuring in Kerr/CFT by taking a further 
decoupling limit (going to `very near horizon region' that has an $AdS_2$) 
on non-chiral $CFT_2$ that corresponds to usual $AdS_3$. One of the 
Virasoro algebras of non-chiral $CFT_2$ becomes Virasoro  of chiral 
$CFT_2$ occuring in Kerr/CFT. One should keep in mind, though, that 
the whole chiral $CFT_2$ does not live in the very near horizon geometry 
($U(1)$ fibred $AdS_2$ throat structure) at fixed $P$ because representation 
of Virasoro algebra includes states with different momentum. Because of the 
extremality constraint in Kerr/CFT, one can say that Virasoro algebra 
contains states above extremal limit but we only consider extremal states. 
So one can still compute entropy of extremal black holes from the Virasoro 
algebra. Since in  \cite{Castro:2008ms}, the authors got chiral $CFT_2$ 
corresponding to $AdS_2$ by dimensional reduction from non-chiral $AdS_3$ CFT, 
we can make a link between two chiral CFT's obtained from non-chiral $AdS_3$
(though, see, the footnote 5).

Therefore, it is very tempting to interpret 
our result in the context of the entropy function formalism as a 
central charge in $AdS_2$. In this way, one can obtain a concrete relation 
between the Kerr/CFT correspondence and the $AdS_2/CFT_1$ duality. However, 
our proposal should be taken with caution: one should explicitly check 
that the boundary conditions imposed in three dimensions are directly 
related to the ones in two dimensions. We leave a more detailed 
analysis of the central charge in this context for future work.
\\
\\
{\bf Acknowledgement}
\\
We thank Andres Anabalon, Ashoke Sen, and Stefan Theisen for valuable discussions, 
Geoffrey Compere for important comments on an earlier draft, and Tatsuo Azeyanagi 
for correspondence. We are also grateful to the organizers of the ISM08 conference 
for a stimulating environment. DA would like to thank Harish-Chandra Institute for 
hospitality during initial stages of  this work and NSERC of Canada for support. 
\\
\\
{\bf Note added} 
\\
While this paper was being completed, refs.\cite{Compere:2009dp, Hotta:2009bm} 
appeared that are related with the present work. In \cite{Compere:2009dp} 
it was also proposed that the Kerr/CFT correspondence can be applied 
to a general class of extremal black hole solutions. In \cite{Hotta:2009bm}
it was also pointed out a possible connection between Kerr/CFT correspondence 
and the attractor mechanism. 

\appendix
\section{Appendix}
\renewcommand{\theequation}{A.\arabic{equation}}
\setcounter{equation}{0}

In this appendix, we present details about our calculation of 
applying Kerr/CFT analysis to GR black hole. As expected in five dimensions, 
we have two $U(1)$'s, corresponding to two azimuthal angles and hence 
one can have two boundary conditions. 

One of the possible boundary conditions for $h_{\mu\nu}$ is
\begin{align}
\left(
\begin{array}{ccccc}
h_{\tau\tau}=\mathcal{O}(r^2)
&h_{\tau r}=\mathcal{O}(\frac{1}{r^2})
&h_{\tau \theta}=\mathcal{O}(\frac{1}{r})
&h_{t\psi}=\mathcal{O}(r)
&h_{t\phi}=\mathcal{O}(1)\\
h_{r\tau}=h_{\tau r}
&h_{rr}=\mathcal{O}(\frac{1}{r^3})
&h_{r\theta}=\mathcal{O}(\frac{1}{r^2})
&h_{r\psi}=\mathcal{O}(\frac{1}{r^3})
&h_{r\phi}=\mathcal{O}(\frac{1}{r^2}) \\
h_{\theta \tau}=h_{\tau\theta}
&h_{\theta r}=h_{r\theta}
&h_{\theta\theta}=\mathcal{O}(\frac{1}{r})
&h_{\theta\psi}=\mathcal{O}(\frac{1}{r})
&h_{\theta \phi} = \mathcal{O}(\frac{1}{r}) \\
h_{\psi \tau}=h_{\tau \psi}
&h_{\psi r} =h_{r\psi}
&h_{\phi\theta} =h_{\theta\phi}
&h_{\psi\psi}=\mathcal{O}(\frac{1}{r})
&h_{\psi \phi}=\mathcal{O}(1)\\
h_{\phi \tau}=h_{\tau \phi}
&h_{\phi r}= h_{r\phi}
&h_{\phi \theta}=h_{\theta \phi}
&h_{\psi\phi} =h_{\phi \psi}
&h_{\phi\phi}=\mathcal{O}(1)
\end{array}
\right),
%\label{ry_boundary_condition}
\end{align}
Let us now find the most general diffeomorphism that preserves the boundary 
conditions --- we have to evaluate the Lie derivatives of $g_{\mu\nu}$
with respect to the vector fields $\zeta$ that preserve the asymptotic 
symmetries:
\bea
{\cal L}_\zeta g_{\mu\nu}=
  \zeta^\rho\partial_\rho g_{\mu\nu} + g_{\rho\nu}\, \partial_\mu \zeta^\rho +
g_{\mu\rho}\, \partial_\nu \zeta^\rho 
\eea 
We obtain (by keeping just the terms which have a non-trivial contribution): 
\bea
h_{\tau\tau}={\cal L}_{\zeta} g_{\tau\tau}\simeq
\zeta^r\partial_rg_{\tau\tau} \simeq O(r^2) 
\Rightarrow \zeta^r=rF(\theta,\psi,\phi)+O(1)
\eea
\bea
h_{\theta\theta}={\cal L}_\zeta g_{\theta\theta}\simeq
g_{\theta\theta}\partial_{\theta}\zeta^{\theta} \simeq O(\frac{1}{r}) 
\Rightarrow \zeta^{\theta}=O(\frac{1}{r})
\eea
\bea
h_{\theta\tau} &=& {\cal L}_\zeta g_{\theta\tau}\simeq
g_{\rho\tau}\partial_{\theta}\zeta^{\rho}=
g_{\phi\tau}\partial_{\theta}\zeta^{\phi}+
g_{\psi\tau}\partial_{\theta}\zeta^{\psi}+
g_{r\tau}\partial_{\theta}\zeta^{r}
\simeq O(\frac{1}{r})\nn 
&\Rightarrow& \zeta^{\phi}=G(\theta,\psi,\phi)+O(\frac{1}{r^2}), \,\, 
\zeta^{\psi}=H(\theta,\psi,\phi)+O(\frac{1}{r^2})\nn
&&g_{\phi\tau}\partial_{\theta}G + 
g_{\psi\tau}\partial_{\theta}H + rg_{r\tau}\partial_{\theta}F=0
\eea
\bea
h_{\theta\phi} &=& {\cal L}_\zeta g_{\theta\phi}\simeq
g_{r\phi}\partial_{\theta}\zeta^{r}+
g_{\phi\phi}\partial_{\theta}\zeta^{\phi}+
g_{\phi\psi}\partial_{\theta}\zeta^{\psi}
\simeq O(\frac{1}{r})\nn 
&& g_{\phi\phi}\partial_{\theta}G + 
g_{\phi\psi}\partial_{\theta}H + 
rg_{r\phi}\partial_{\theta}F=0
\eea
\bea
h_{\theta\psi} &=& {\cal L}_\zeta g_{\theta\psi}\simeq
g_{r\psi}\partial_{\theta}\zeta^{r}+
g_{\psi\psi}\partial_{\theta}\zeta^{\psi}+
g_{\phi\psi}\partial_{\theta}\zeta^{\phi}
\simeq O(\frac{1}{r})\nn 
&& g_{\phi\psi}\partial_{\theta}G + g_{\psi\psi}\partial_{\theta}H + rg_{r\psi}\partial_{\theta}F=0
\eea
Up to this point we obtained $G(\psi,\phi), H(\psi,\phi),$ and 
$F(\psi,\phi)$. The next relation removes the dependence of $\psi$:
\bea
h_{\psi\psi} &=& {\cal L}_\zeta g_{\psi\psi}=
\zeta^{\theta}\partial_{\theta}g_{\psi\psi}+
2g_{r\psi}\partial_{\psi}\zeta^{r}+
2g_{\psi\psi}\partial_{\psi}\zeta^{\psi}+
2g_{\phi\psi}\partial_{\psi}\zeta^{\phi}
\simeq O(\frac{1}{r})\nn 
&& \partial_{\psi} \left[g_{\psi\psi}H+
g_{\phi\psi}G+rg_{r\psi}F\right]=0
\eea
In fact even this one supports the non-dependence of $\psi$ and also imposes a 
constraint on $\zeta^{\tau}$:
\bea
h_{r\psi} &=& g_{\tau\psi}\partial_r\zeta^{\tau}+\partial_{\psi}\left[\
g_{rr}\zeta^r+g_{r\psi}\zeta^{\psi}+g_{r\phi}\zeta^{\phi}\right]
\simeq O(\frac{1}{r^3})\nn
&& \zeta^{\tau}=C+O(\frac{1}{r^3})
\eea
We also obtain 
\bea
h_{\tau\phi} &=& \zeta^{r}\partial_{r}g_{\tau\phi} + g_{\tau\tau}\partial_{\phi}\zeta^{\tau}+g_{\tau\phi}\partial_{\phi}\zeta^{\phi} + g_{r\tau}\partial_{\phi}
\zeta^{r} + g_{\tau\psi}\partial_{\phi}\zeta^{\psi} \nn
&& h_{\tau\phi} = \zeta^{r}\partial_{r}g_{\tau\phi} + g_{\tau\phi}\partial_{\phi}\zeta^{\phi} + O(1) + 
g_{r\tau}\partial_{\phi}\zeta^{r}
\eea
The first two terms cancel giving us the required $F+G'=0$ relation and $O(1)$ 
term matches the $O(1)$ of $h_{\tau\phi}$. But the term $g_{r\tau}\partial_{\phi}\zeta^{r}$ gives an $O(r)$ contribution because $\zeta^{r}$ is $O(r)$. One can try to change 
$\zeta^{r}$ but that conflicts with other equations. So one must 
set $g_{r\tau}=0$ to avoid this problem. For the case where we set it to zero, 
we have the usual vector fields which give the central charge. One can 
always perform coordinate transformation to get rid of $g_{r\tau}$ term 
or after dimensional reduction to four dimensions, one can perform a gauge 
transformation to get rid of this component of the gauge field.

So finally we get the result that the most general diffeomorphism that preserves 
the boundary condition is given by
\begin{align}
\zeta &=
\Bigl[C+\mathcal{O}\bigl(\frac{1}{r^3}\bigr)\Bigr]\partial_t
+[-r\gamma'(\phi)+\mathcal{O}(1)]\partial_{r}
+\mathcal{O}\bigl(\frac{1}{r}\bigr)\partial_{\theta} \no
&\quad +\mathcal{O}\bigl(\frac{1}{r^2}\bigr)\partial_{\psi}
+\Bigl[\gamma(\phi)+\mathcal{O}\bigl(\frac{1}{r^2}\bigr)\Bigr]\partial_\phi,
\end{align}
where $C$ is an arbitrary constant and
$\gamma(\phi)$ is an arbitrary function of $\phi$.
From this, the asymptotic symmetry group is generated by 
the diffeomorphisms of the form 
\begin{align}
\zeta^t&=\partial_{t},\\
\zeta^\phi_\gamma&=\gamma(\phi)\partial_{\phi}-r\gamma'(\phi)\partial_r 
\end{align}

\end{document}